\documentclass[10pt, conference]{IEEEtran}
\IEEEoverridecommandlockouts

\ifCLASSINFOpdf
  \usepackage[pdftex]{graphicx}
  \DeclareGraphicsExtensions{.pdf,.jpeg,.png}
\else
  \usepackage[dvips]{graphicx}
  \DeclareGraphicsExtensions{.eps}
\fi
\usepackage[cmex10]{amsmath}
\usepackage{algorithmic}
\usepackage{array}
\usepackage{mdwmath}
\usepackage{mdwtab}
\usepackage{eqparbox}
\usepackage[tight,footnotesize]{subfigure}
\usepackage[caption=false,font=footnotesize]{subfig}
\usepackage{fixltx2e}
\usepackage{stfloats}
\usepackage{url}
\usepackage[american]{babel}

\begin{document}

\title{Geosocial Graph-Based Community Detection}

\author{\IEEEauthorblockN{Yves van Gennip, Huiyi Hu, Blake Hunter}
\IEEEauthorblockA{Department of Mathematics\\
University of California, Los Angeles\\
Los Angeles, CA, USA\\
Email: yvgennip@math.ucla.edu, huiyihu@math.ucla.edu,\\ blakehunter@math.ucla.edu}
\and
\IEEEauthorblockN{Mason A. Porter}
\IEEEauthorblockA{Oxford Centre for Industrial and Applied Mathematics,\\ Mathematical Institute;\\ and CABDyN Complexity Centre
\\
University of Oxford\\
Oxford, UK\\
Email: porterm@maths.ox.ac.uk}
}

\maketitle

\begin{abstract}

We apply spectral clustering and multislice modularity optimization to a Los Angeles Police Department field interview card data set. To detect communities (i.e., cohesive groups of vertices), we use both geographic and social information about stops involving street gang members in the LAPD district of Hollenbeck. We then compare the algorithmically detected communities with known gang identifications and argue that discrepancies are due to sparsity of social connections in the data as well as complex underlying sociological factors that blur distinctions between communities. 
\end{abstract}



\section{Introduction}

\begin{figure}[ht]
	\begin{center}
	\includegraphics[width=.45\textwidth]{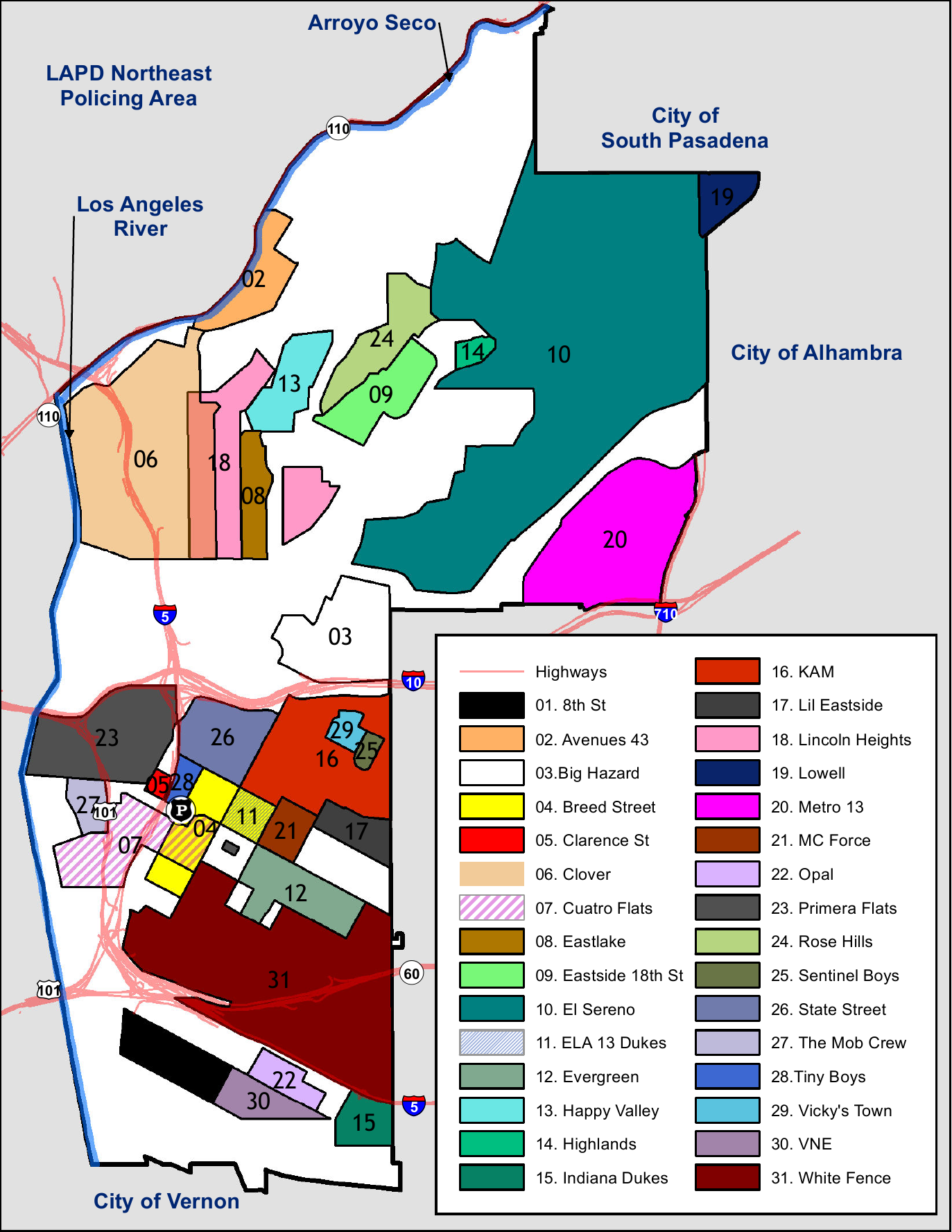} 
	\end{center}
	\caption {\footnotesize Map of gang territories in the Hollenbeck area of Los Angeles. [Figure courtesy of Matthew Valasik, UC Irvine.]
	  }\label{fig:Hollenbeckdata}
\end{figure}

Many networks can be partitioned into \emph{communities}, such that they consist of cohesive (and often dense) groups of vertices with sparse connections between distinct groups \cite{comnotices}. In this paper, we algorithmically detect communities in a social network based on sparse geosocial information.  The data come from the policing district Hollenbeck (see Fig.~\ref{fig:Hollenbeckdata}) in Los Angeles and were collected using Field Interview cards (FI cards) from 2009, which the Los Angeles Police Department (LAPD) collected when interacting with the public. The vast majority of these stops are noncriminal, and the data include both the location of the stops and the individuals involved in them. Using this information, we perform unsupervised clustering on 748 known gang members to produce groups that we subsequently compare with known gang affiliations. We consider two graph-based community-detection techniques: spectral clustering and multislice modularity optimization.

\section{Methods}


To apply spectral clustering or modularity optimization, we represent the data as a graph.  We construct the graph using a normalized adjacency matrix $D^{-1} W$, where the matrix $W$ has edge weights
\begin{equation}\label{weights} 
	W_{i,j} = \alpha S_{i,j} + (1-\alpha) e^{-d^2_{i,j} / \sigma^2}\,,
\end{equation}	
and $D$ is a diagonal matrix whose $i$th nonzero entry $D_{i, i} = \sum_j W_{i,j}$ is the strength (i.e., weighted degree) of vertex $i$. The matrix $S$ captures social interactions of individuals: $S_{i,j}=1$ if $i$ and $j$ met and $S_{i,j} = 0$ otherwise. The second term in (\ref{weights}) uses the Euclidean distance $d_{i,j}$ between the mean locations of individuals $i$ and $j$ to describe geographic similarity. The scale parameter $\sigma$ is chosen to be the length which is one standard deviation larger than the mean distance between two individuals who have been stopped together (but most results are fairly robust to small changes in its value). The parameter $\alpha \in [0,1]$ allows us to control the relative contributions of the social and geographic information in the construction of the graph. 

Spectral clustering \cite{ShiMalik00,NgJordanWeiss02} uses the eigenvectors of a graph's adjacency matrix to cluster the vertices. The first $k$ eigenvectors of $D^{-1}W$ are approximate indicator functions for the clustering that solves the NP-complete normalized cut minimization problem \cite{ShiMalik00,VonLuxburg07,YuShi03}. Consequently, each data point is given coordinates matching the corresponding entries of the first $k$ eigenvectors of the normalized geosocial adjacency matrix $D^{-1}W$. Thus the $j^\text{th}$ data point (out of $n$ points) is given coordinates $(v^1_j, \ldots, v^k_j)$, where the eigenvectors are $v^i$, $i, j=1,\ldots, k$, with entries $v^i_j$, $j=1, \ldots n$. The network can subsequently be partitioned using $k$-means clustering \cite{HartiganWong79} on the new coordinates. Because the data set contains members of 31 gangs, we prescribe 31 clusters for the $k$-means algorithm.  The geo-social data presented in this paper is high-dimensional and contains a complex overlapping cluster structure.  The complex nature of the data, the simplicity of the method's construction, and the existence of efficient solvers makes spectral clustering appropriate for this problem.


Modularity optimization \cite{comnotices,NewmanGirvan04} is a community-detection method that does not require prior knowledge of the number of desired communities. It finds cohesive groups within a network by comparing the network with a null model.  One seeks a partition that maximizes the quality function
\begin{equation}\label{modularity}
	Q = \frac{1}{2m}\sum_{i,j} \left[(D^{-1}W)_{i,j} - \gamma P_{i,j}\right]\delta_{i,j}\,,
\end{equation}	
which measures the aggregate strength of edges within communities compared to the aggregate strength obtained using a random null-model network with entries $P_{i,j}$.  We use $P_{i,j}=\frac{d_i d_j}{\sum_i d_i}$, which preserves the network's expected strength distribution but otherwise randomizes the data.  We use a resolution parameter $\gamma$ to examine communities at multiple scales \cite{ReichardtBornholdt06,FortunatoBarthelemy07} (the canonical value for modularity optimization is $\gamma = 1$), and we use the Kronecker delta $\delta_{i,j}$ to indicate the event that vertices $i$ and $j$ belong to the same community.  By maximizing the modularity $Q$ in (\ref{modularity}), which we do using a (locally greedy) Louvain method \cite{Blondel2008}, we aim to algorithmically detect communities with significantly stronger intra-community connections than expected by chance.  
Modularity was generalized by Mucha et. al \cite{MuchaRichardsonMaconPorterOnnela10} to 
``multislice" networks, which consist of layers of ordinary networks in which vertex $i$ in one slice is connected to the corresponding vertex in other slices, via a coupling constant $\omega$.  For our example, each slice is a copy of the graph in which we wish to consider a different value of the resolution parameter $\gamma$, and we connect each vertex in every slice to the corresponding vertex in neighboring slices (with the slices ordered according to the values of $\gamma$) using interslice edges. This allows us to detect communities simultaneously over a range of resolution parameters, while still enforcing some consistency in clustering identical vertices similarly across slices.  We subsequently examine the output of multislice modularity optimization using network diagnostics such as the number of clusters, purity, and $z$-scores of Rand coefficients. This application of multislice modularity optimization to the FI card data set is an extension of what was done in \cite{vanGennipHunterAhnElliotLuhHalvorsonReidValasikWoTitaBertozziBrantingham12}.

We investigate whether the geosocial information from the LAPD FI card data suffices to detect community structures that match the LAPD's notion of the gang communities in Hollenbeck. Both methods perform well in finding communities, but we find that two factors conspire to make some clusters inhomogeneous in their gang composition. First, the social connections between gang members are very sparse in the data. Second, sociological reality suggests that gang boundaries are not as rigorous as is often believed. Just like other people, gang members are known to play multiple social roles (only one of which is being a fellow gang member) \cite{Pattillo03}, as they can also be fathers, sisters, colleagues, teammates, etc.  Hence, although 88.7\% (423 out of 477) of all social connections in our data are intra-gang contacts, inter-gang contact is not a rare occurrence.


\section{Results}

\begin{figure}[ht]
    \begin{center}
    $\begin{array}{c}
    \includegraphics[width=0.5\textwidth]{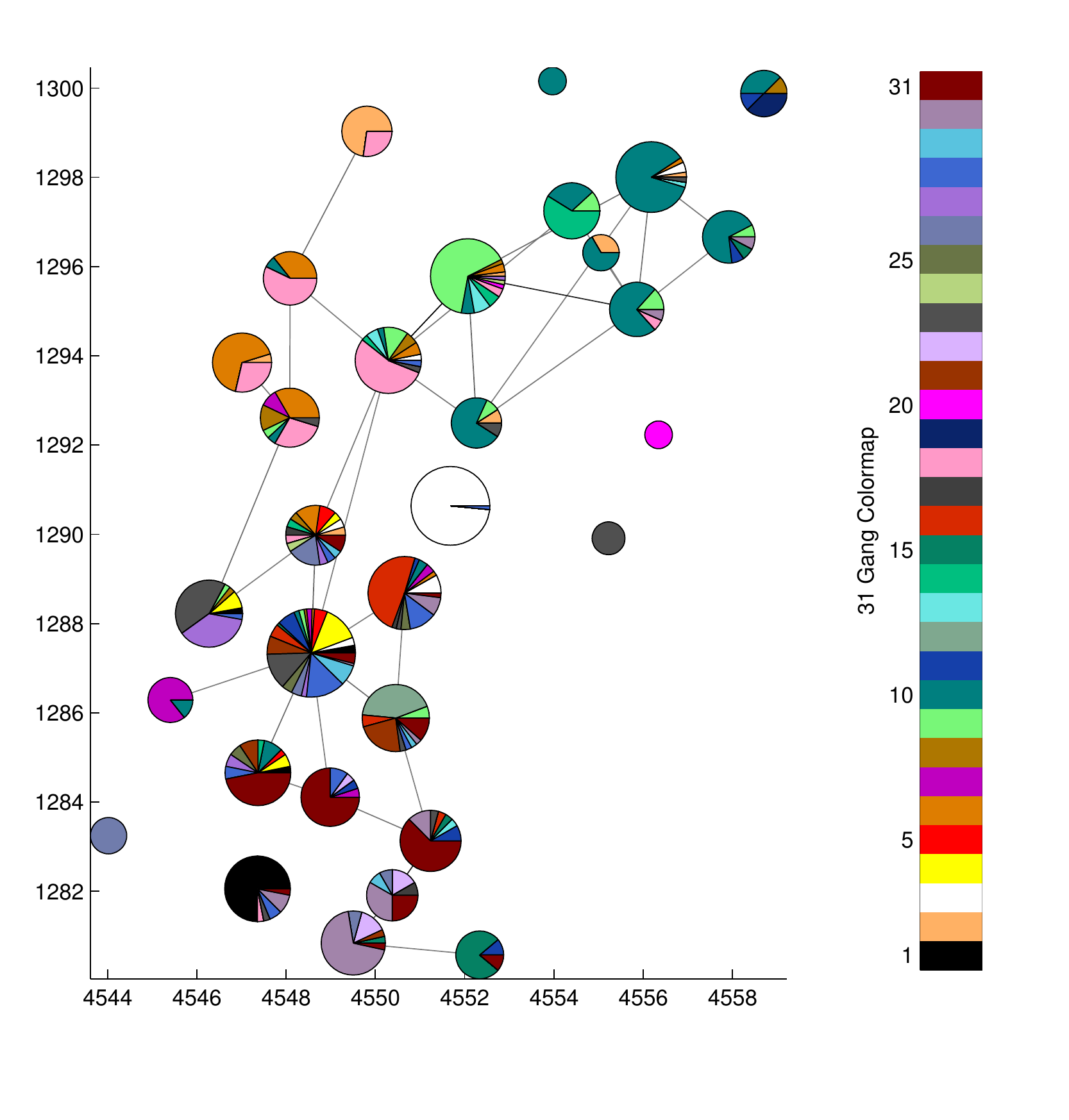}
    \end{array}$
    \end{center}
\caption{\footnotesize Pie charts made using the network visualization code from Ref.~\cite{TraudFrostMuchaPorter09} for spectral clustering and $\alpha=0.4$.  Each pie represents a community, and its size indicates the number of gang members in it.
The coloring, which we obtain from the gang colors in Fig.~\ref{fig:Hollenbeckdata}, indicates the gang composition of the communities (see the legend). Each pie is centered at the centroid of the mean locations of the individuals in its corresponding cluster. The numbering in the axes uses an arbitrary but fixed origin. For aesthetic reasons, the units on each axes are approximately $435.42$ meters. The connections drawn between different pies indicate inter-community social connections (i.e., nonzero elements of $S$).
}\label{fig:piechartclustering}
\end{figure}

We compare the partition into gangs provided by the LAPD with the communities that we obtained using spectral clustering and multislice modularity optimization using purity and $z$-Rand scores. To compute purity, see for example \cite{HarrisAubertHaeb-UmbachBeyerlein1999}, we assign to all vertices in a given community the label of the gang that appears the most often in that group (in case of a tie between two or more gangs, the label of one of these gangs is arbitrarily chosen for all the vertices in that group). The purity is then given by the fraction of the correctly labeled individuals in the whole data set. To obtain the $z$-Rand score, we compute the number of pairs 
$w$ of individuals who belong to the same gang and who are placed in the same community by a clustering algorithm. We then compare this number to its expected value under a hypergeometric distribution with the same number and sizes of communities. The $z$-Rand score, which is normalized by the standard deviation from the mean, indicates how far the actual $w$ value lies in the tail of a distribution \cite{TraudKelsicMuchaPorter11}. 

For spectral clustering, we compute the mean purity and $z$-Rand scores, with error margins given by the standard deviations, over 10 runs of $k$-means clustering. The mean purity is about 0.55 within error margins independent of $\alpha$, unless $\alpha=1$ (i.e., when we only use social information), for which the purity plummets to about 0.25. We see a similar trend in the $z$-Rand scores, which fluctuate within error margins around 140 for all values of $\alpha$ except near $\alpha=1$, for which $z$-Rand is about $6$. In Fig.~\ref{fig:piechartclustering} we show the resulting clustering from a run of $k$-means clustering with $\alpha=0.4$.  Note that the $z$-Rand score of the partition into true gangs is about 405.


\begin{figure}[h]
\includegraphics[width=0.50\textwidth]{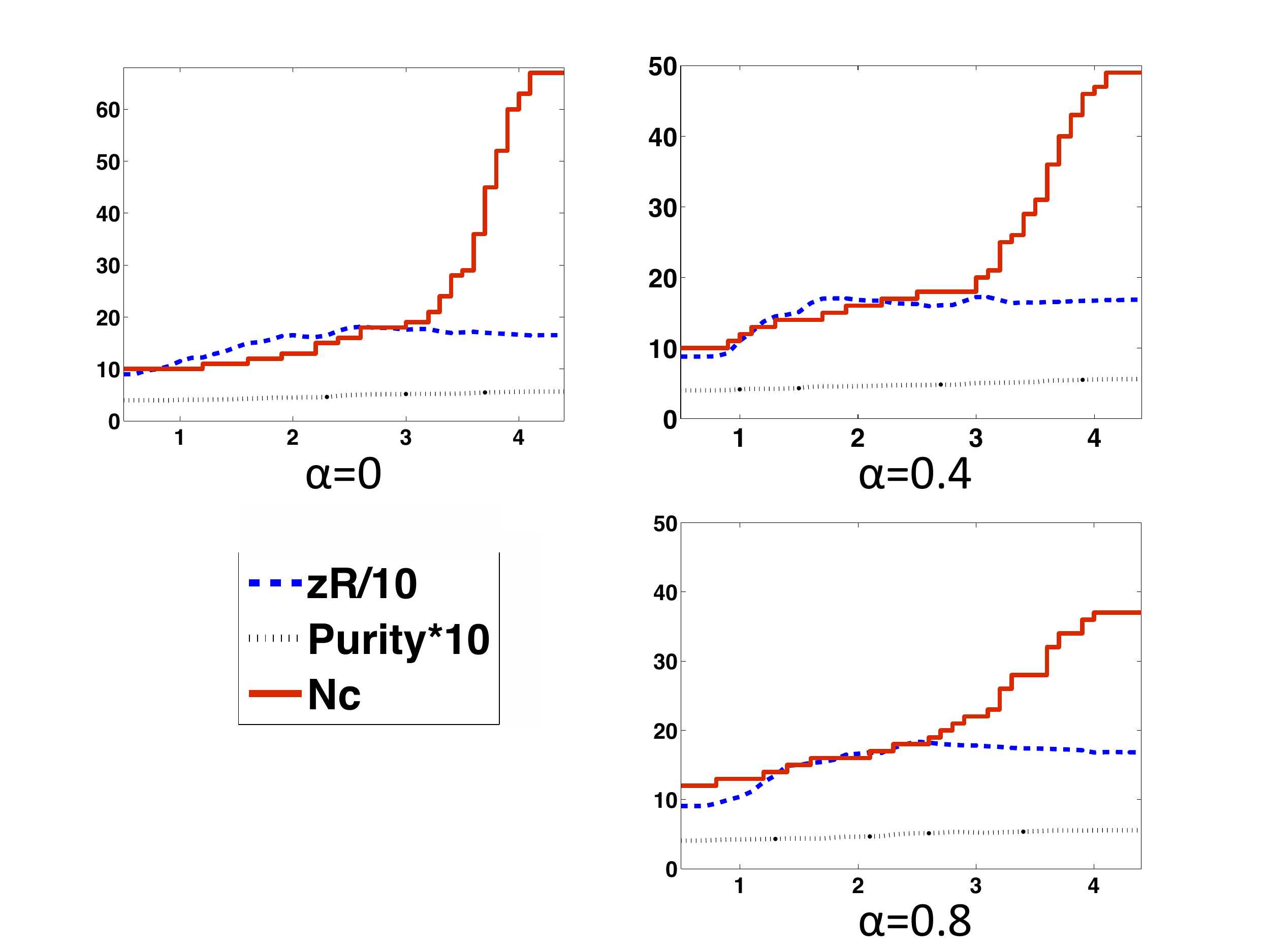}
 \caption{Number of communities, purity, and $z$-Rand scores (vertical axes) as a function of resolution parameter $\gamma$ (horizontal axes) for three values of $\alpha$ and an interslice coupling constant of $\omega =1$.}\label{fig:multiplex}
\end{figure}

Multislice modularity optimization determines the number of clusters as part of its output. We plot this number, together with the purity and $z$-Rand scores for the resulting network partitions, as a function of the resolution parameter and for different values of $\alpha$ in Fig.~\ref{fig:multiplex}. We use an interslice coupling constant of $\omega=1$. Small variations in the value of $\omega$ did not qualitatively change the result. We seek plateaus in the number of clusters that are near a local maximum of the $z$-Rand score. The details differ slightly for different $\alpha$, but the general picture that arises is that the optimal number of clusters for our data lies around 18 clusters with a resulting $z$-Rand score of about 180. Purity is again roughly constant and again near 0.5. Note, however, that comparing the purity scores of two different partitions with different numbers of clusters is not very meaningful, as purity is biased to favor partitions with more communities.



For both of the clustering methods that we considered, changing the value of $\alpha$ (as long as it is strictly less than $1$) does not have a big influence on the resulting purity and $z$-Rand scores. This suggests that the social component of our data set is too sparse to significantly improve algorithmic clustering. To test whether an improvement might be expected at all if more information on social interactions becomes available in the future, we construct a \textit{ground-truth derived social matrix} $GT(p,q)$ and run the spectral clustering algorithm using $GT(p,q)$ as our social matrix (and the same geographical matrix as before).

The matrix $GT(p,q)$ contains a fraction $p$ of the possible intra-gang connections (true positives), a fraction $q$ of which we change from true positives to false positives to simulate noise. In a sense, $p$ indicates how many connections are observed and $q$ can be construed as approximating how many of those are between members of different gangs. The matrix $GT(1, 0)$ is the full intra-gang matrix; it contains nonzero entries (of value $1$) only for each pair of individuals from the same gang. Sampling a fraction $p$ of these connections (uniformly at random) from the strictly upper triangular part of the matrix, setting the others to $0$, and then symmetrizing the matrix, gives $GT(p,0)$. Finally, in $GT(p,q)$, we set a fraction $q$ (again sampled uniformly at random) of all of the nonzero entries from the strictly upper triangular part to $0$, and we set the same number of $0$ entries to $1$, and symmetrize again. In this process, we preserve the diagonal entries at $1$ and the symmetry of the matrix.



\begin{figure}[ht]
	\begin{center}
		\includegraphics[width=0.5\textwidth]{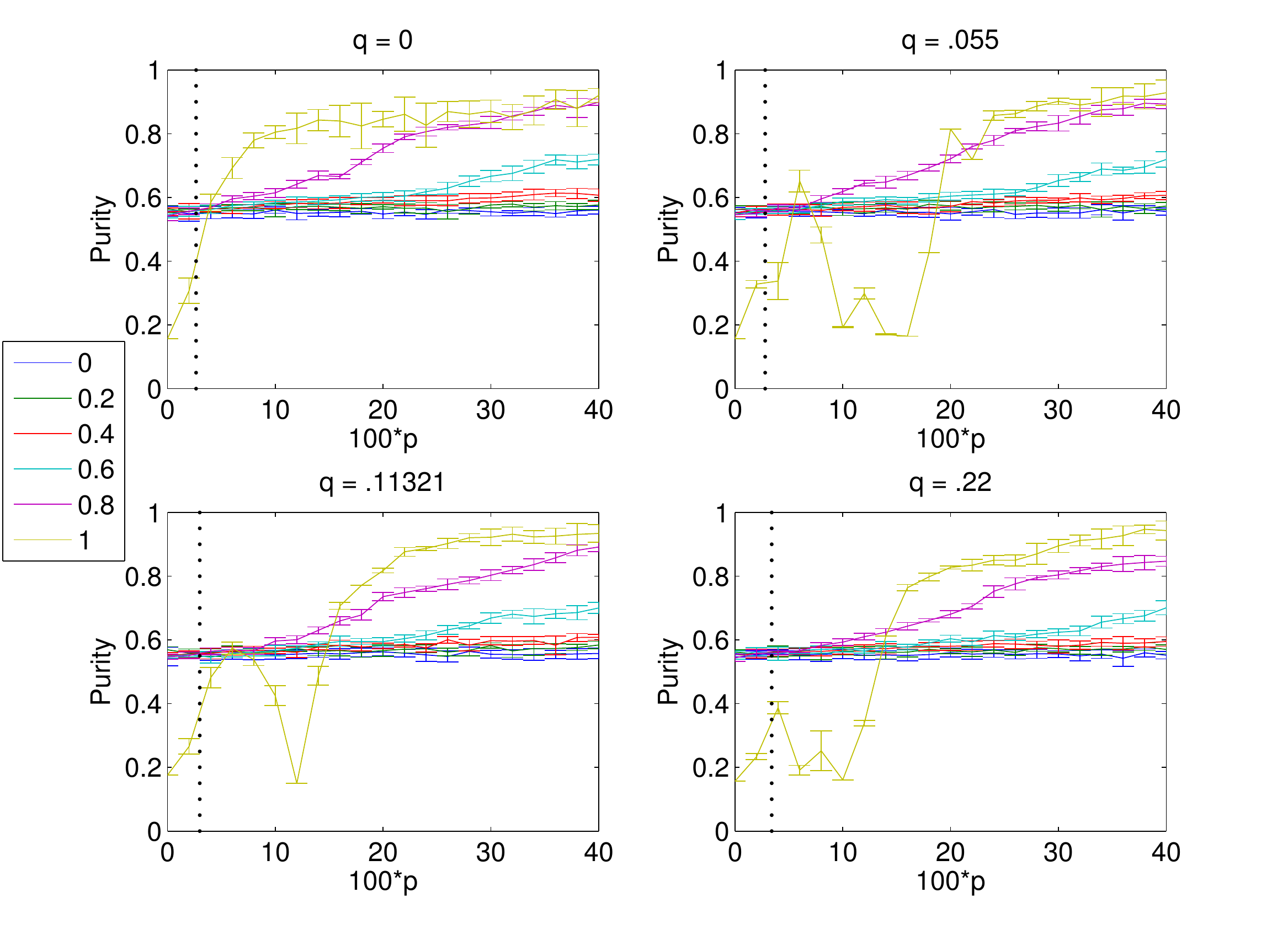}\\
	\end{center}
	\caption{\footnotesize Plots of the purity using $S=GT(p,q)$ in the spectral clustering algorithm for different values of $q$ (the different plots) and $\alpha$ (the different curves in each plot) as a function of $p$.  For each set of parameter values, we compute a purity score from an average over 10 runs of $k$-means clustering, and the error bars give the standard deviation over these runs. The dotted vertical lines indicate the values of $p$ for which the number of true positives in $GT(p,q)$ is equal to the number of true positives in $S$.
	}\label{fig:varyingpq}
\end{figure}

In Fig.~\ref{fig:varyingpq}, we show the mean purity scores over 10 runs of $k$-means clustering as a function of $p$ for various values of $q$. The empirical values of $p$ and $q$ for our data set are $p\approx 0.266$ and $q\approx 0.1132$ (see the dotted line in the lower left panel of the figure). We see that increasing $p$, which can be interpreted as collecting more field data about the intra-gang social connections, has a bigger impact on purity than lowering $q$. Using these dotted lines as guidance, we also see that we cannot expect our current data set to yield network partitions of significantly higher purity than those we obtained.


We also apply a simple Gaussian mixture model and $k$-means algorithm to compare with our spectral clustering and multislice modularity results. 

For the Gaussian mixture model \cite{McLachlanPeel00}, we run the MATLAB function {\tt gmdistribution.fit} on the mean location vectors of the individuals in our data set.  (It was not possible to include the social data in the simple model that we used.) This function uses an expectation maximization (EM) algorithm to fit 31 two-dimensional Gaussians to the data. This results in a set of 31 means and 31 standard deviations. To divide the individuals over 31 different clusters, we assign each point to its nearest mean using a normalized distance. (We normalized the distance to each mean by the corresponding variance.)  Using this technique, we again find a purity of about 0.55, which again demonstrates that geography alone does indeed account for most of the purity. The $z$-Rand score is about 100, which is worse than the results that we get with spectral clustering. Clearly, the inclusion of social data (even sparse social data) improves the $z$-Rand score. In this context, it is noteworthy that the $z$-Rand score for spectral clustering for $\alpha=0$ is about 120, with a standard deviation of about $19$. Hence, the $z$-Rand score for the Gaussian mixture model is only slightly more than one standard deviation removed from the $\alpha=0$ spectral clustering score. It is not trivial to include the social data in the Gaussian mixture model, so spectral clustering seems preferable for situations like the present that combine different types of data.

To implement $k$-means directly, we run the MATLAB function {\tt kmeans} on the columns of the matrix $D^{-1} W$. We compute averages over 10 $k$-means runs, and we find mean purity scores of about 0.56, which is again comparable (within about one standard deviation) to the spectral clustering results, for all $\alpha$ up to about $\alpha = 0.8$. Among these $\alpha$-values, the mean $z$-Rand scores vary quite a bit, but again they typically lie within one standard deviation (usually with a somewhat higher value) of their spectral clustering counterpart. Interestingly, however, for larger values of $\alpha$, $k$-means clustering by itself performs quite a bit worse than spectral clustering (using the same $\alpha$ values). Clearly, the embedding using the eigenvectors of $D^{-1} W$ which spectral clustering uses, is needed to make the complicated geosocial data structure amenable to $k$-means clustering. Our results from Fig.~\ref{fig:varyingpq} demonstrate that, while for the current sparse social data the added benefit of incorporating this data into our method is limited, if the social data were slightly less sparse (i.e., if the value of $p$ were higher), there would be a clear benefit from including it, especially when $\alpha$ is large. The results using the $k$-means algorithm by itself show that, exactly for these high $\alpha$ values, $k$-means performs worse than spectral clustering, suggesting that spectral clustering is preferable to $k$-means clustering by itself, if less sparse social data become available. Because optimization of multislice modularity performed at a comparable level to spectral clustering on the data that we studied, we can also draw similar conclusions when comparing Gaussian mixture model, $k$-means algorithm, and optimization of multislice modularity.


\section{Conclusions}

We study communities among gang members in the LAPD division Hollenbeck by applying spectral clustering and multislice modularity optimization to an LAPD FI card data set from 2009. Using only information about where and with whom the gang members were stopped by the police, we partition a network representation of this data into communities that correspond to their actual gang affiliations with a purity of about 0.5.  We demonstrated, however, that this lack of purity seems to arise from the sparsity of intra-gang connections in the data. It is an interesting question whether this sparsity is due to the data collecting methods---that is, whether or not there are many additional unrecorded and/or unobserved intra-gang social interactions in public---or whether it is an inherent property of the system (e.g., perhaps members of the same gang do not interact with each other particularly often in public). The additional fact that each individual in the data set is on average connected to only 1.2754 $\pm$ 1.8946 (with the number always nonnegative, of course) other people suggests that the former explanation might play a dominant role.
Indeed, the maximum number of connections for an individual in the data set is 23, but 315 of the 748 gang members (42\%) are not connected to any other individual, and it seems all but inconceivable that such a high percentage of gang members truly never interact with any other gang members in public.

It has been documented that gangs can vary substantially in their extent of internal organization \cite{DeckerCurry02a}. The large mixed-gang community that we observe in Fig.~\ref{fig:piechartclustering} near the coordinates (4549, 1287) is located in an area of a housing project where several gangs claimed turf.  At the time the data were collected, this project had been recently reconstructed and had displaced resident gang members.  However, even with these individuals scattered across the city, they seemed to remain tethered to their social spaces in their established territories. \cite{HUD98,Olivo01}

Further studies of mathematical, data analytical, and sociological nature will hopefully shed additional light on the question whether gangs are sharply delineated social groups. In the current study, we have attempted to illuminate one piece of this puzzle. 


%
%


\section*{Acknowledgments}

We thank the Hollenbeck Division of the LAPD and Megan Halvorson, Shannon Reid, Matthew Valasik, James Wo, and George E. Tita, at the Department of Criminology, Law, and Society of UCI, for collecting, digitizing, and cleaning the data. We are grateful to P. Jeffrey Brantingham for many useful discussions.

This paper arose initially from an undergraduate project by Raymond Ahn, Peter Elliott, and Kyle Luh, that was mentored by the first and third author during the 2011 UCLA Applied Mathematics REU. This program was organized by Andrea L. Bertozzi and funded by NSF grant DMS-1045536. Further study was made possible by funding from NSF grant DMS-0968309, AFOSR MURI grant FA9550-10-1-0569 and ONR grants N000141010221, N000141210040, and N000141210838. MAP thanks Andrea Bertozzi for hosting his visit to UCLA, and he acknowledges a research award (\#220020177) from the James S. McDonnell Foundation.


\bibliographystyle{IEEEtran}
\bibliography{IEEEabrv,bibliography,mason}


\end{document}